\documentclass[11pt]{article}

\usepackage{silence}
\usepackage{amsfonts}
\usepackage[margin=1in]{geometry}
\usepackage{subcaption}
\usepackage{amssymb,amsmath}
\usepackage{bbm}
\usepackage{lmodern}
\usepackage{graphicx}
\graphicspath{{./figs_arxiv}}
\usepackage[outdir=./figs]{epstopdf}
\usepackage{algorithm}
\usepackage{algpseudocode}
\usepackage{hyperref}
\usepackage[capitalize]{cleveref}
\usepackage{resizegather}
\usepackage{enumitem}

\usepackage{color}

\crefformat{equation}{(#2#1#3)}

\usepackage{amsopn}
\newcommand{\mycomment}[1]{}

\DeclareMathOperator{\e}{\text{e}}
\DeclareMathOperator{\R}{\mathbb{R}}

\def\L{\mathcal L}
\def\rK{\reflectbox{K}}
\def\rKy{\widehat{\rK}}
\def\K{K}

\newtheorem{remark}{Remark}

\usepackage{subfiles} 

\begin{document}

\title{Bayesian computation with generative diffusion models by Multilevel Monte Carlo}

\author{Abdul-Lateef Haji-Ali$^{1,4}$, Marcelo Pereyra$^{1,4}$, Luke Shaw$^{2}$, Konstantinos Zygalakis$^{3,4}$\\[2ex]
$^{1}${\small\it School of Mathematical and Computer Sciences, Heriot-Watt University, Edinburgh, EH14 4AS, UK}\\[1ex]
  $^{2}${\small\it Departament de Matem\`{a}tiques and IMAC, Universitat Jaume I, 12071-Castell\'{o}n de la Plana, Spain}\\{
\small\it email: shaw@uji.es}\\[1ex]
  $^{3}${\small\it School of Mathematics, University of Edinburgh, Edinburgh, EH9 3FD,
  UK}\\[1ex]
\(^{4}\) {\small\it Maxwell Institute for Mathematical Sciences, Edinburgh, UK.}\\[1ex]
}


\maketitle
\begin{abstract}
Generative diffusion models have recently emerged as a powerful strategy to perform stochastic sampling in Bayesian inverse problems, delivering remarkably accurate solutions for a wide range of challenging applications. However, diffusion models often require a large number of neural function evaluations per sample in order to deliver accurate posterior samples. As a result, using diffusion models as stochastic samplers for Monte Carlo integration in Bayesian computation can be highly computationally expensive, {particularly in applications that require a substantial number of Monte Carlo samples for conducting uncertainty quantification analyses}. This cost is especially high in large-scale inverse problems such as computational imaging, which rely on large neural networks that are expensive to evaluate. {With quantitative imaging applications in mind, this paper presents a Multilevel Monte Carlo strategy that significantly reduces the cost of Bayesian computation with diffusion models}. This is achieved by exploiting cost-accuracy trade-offs inherent to diffusion models to carefully couple models of different levels of accuracy in a manner that significantly reduces the overall cost of the calculation, without reducing the final accuracy. {The proposed approach achieves a $4\times$-to-$8\times$ reduction in computational cost w.r.t. standard techniques across three benchmark imaging problems.} 
\end{abstract}


\section{Introduction}
Many science and engineering problems require solving an inverse problem that is ill-conditioned or ill-posed \cite{kaipio2006statistical}. Bayesian statistical provides a powerful framework to regularise these problems and deliver meaningful solutions that are well-posed \cite{kaipio2006statistical}. Bayesian analysis is especially useful in {quantitative and scientific} applications that require quantifying the uncertainty in the solutions delivered, so that {said solutions can be used reliably for science and decision-making \cite{RobertBook}}.

Modern Bayesian inversion methods are increasingly strongly reliant on machine learning techniques in order to leverage information that is available in the form of training data \cite{Mukherjee2023}. Deep generative modelling provides a highly effective approach for constructing machine-learning-based Bayesian inversion methods, both for directly modelling posterior distributions \cite{Liu2023,Saharia2022,Saharia2022Palette}, as well as for constructing data-driven priors that can be combined with an explicit likelihood function derived from a physical forward model \cite[Section 5]{Mukherjee2023}. In particular, diffusion models (DMs) have attracted significant attention recently because of their capacity to deliver remarkably accurate inferences. Originally proposed for tasks related to generating and editing creative content \cite{SohlDickstein2015,Song2019, Ho2020, Song2020DDIM}, DMs are now also widely studied for Bayesian inversion, especially in the context of signal processing and computational imaging (see, e.g.,  \cite{Chung2022DPS,Kawar2022,Saharia2022Palette,Liu2023}).

{From a computation viewpoint, DMs are stochastic samplers that rely on neural networks to generate samples from a posterior distribution that is encoded implicitly within the DM. State-of-the-art DMs for Bayesian inversion can produce remarkably accurate samples, but they are also highly computationally costly, as producing each sample requires a large number of evaluations of a large neural network which is expensive to evaluate. Furthermore, while some generative machine learning applications require only a small number of posterior samples to identify representative solutions, quantitative inference tasks often demand the generation of hundreds or thousands of samples to estimate posterior moments and probabilities via Monte Carlo integration. The elevated cost currently constrains the practical implementation of DM-based strategies in such settings, especially when accurate uncertainty quantification is required.}

{Motivated by the needs of generative machine learning applications, efforts to reduce the computational cost of DMs have so far focused predominantly on reducing the cost of each NFE (e.g., pruning \cite{Sunil2022} and quantization \cite{Li2023}), and on reducing the number of NFEs required to generate each sample (e.g., via distillation \cite{Salimans2022} and improved numerical schemes \cite{Song2020DDIM}). In contrast, to the best of our knowledge, the potential for reducing the computational cost of DM-based Bayesian inversion by optimizing NFEs for Monte Carlo integration remains unexplored.} This paper presents a new strategy to significantly reduce the cost of Monte Carlo integration in DM-based Bayesian inversion by leveraging Multilevel Monte Carlo (MLMC) \cite{Giles2008,Giles2015}. {The proposed MLMC method is applicable to any DM, allowing it to be integrated with any of the aforementioned strategies to enhance the computational efficiency of posterior sampling and Bayesian computation through DMs.} 

\section{Proposed Multilevel Monte Carlo approach}
\subsection{Notation and problem statement}\label{sec:notation}
We consider the estimation of an unknown high-dimensional quantity of interest $x\in\R^n$ from some observed data $y\in\R^m$. We formulate the problem in the Bayesian statistical framework and model $x$ as a realisation of a random variable $X$ taking values in $\R^n$, and the observation $y$ as a realisation of an $\R^m$-valued random variable that, conditional on $X=x$, has distribution
\begin{equation}\label{eq:InverseProblemNoisy0}
    Y \sim \mathcal{P}(\mathcal{A}(x)),
\end{equation}
where the operator $\mathcal{A}:\R^n \to \R^m$ encodes deterministic aspects of the observation model, and $\mathcal{P}$ models stochastic aspects such as measurement noise. The forward model \cref{eq:InverseProblemNoisy0} includes, for example, the widely used linear Gaussian observation model
\begin{equation}\label{eq:InverseProblemNoisy}
    y=\mathcal{A}x+\eta.
\end{equation}
where $\mathcal{A} \in \R^{m\times n}$ (often with $m<n$), and the additive perturbation $\eta$ is a realisation of zero-mean Gaussian noise with covariance $\sigma^2\mathbb{I}_n$ for \(\sigma^{2} \in \R_{+}\) and \(\mathbb{I}_{n} \in \R^{n\times n}\) being the identity matrix.

Additional information about the solution is introduced by specifying the prior distribution of $X$, whose density we henceforth denote by $\pi(\cdot)$. Observed and prior information are then combined by using Bayes' theorem to derive the posterior distribution for $X$ given the observation \(Y=y\), with density denoted by \(p(\cdot | y )\) satisfying
\begin{equation}\label{eq:Posterior}
  p(x\vert y) = \frac{\L(y\vert x)\pi(x)}
  {\mathcal E(y)}
  \,,
\end{equation}
for all $x \in \R^n$, where $\L$ is the likelihood function associated to the statistical forward model \cref{eq:InverseProblemNoisy0} and \(\mathcal E(y)\) is the data evidence which is defined so that the posterior density integrates to 1.

As mentioned previously, we consider a purely data-driven scenario in which a DM has been trained to draw (approximate) posterior samples. This could be a DM trained on a dataset $\{x^{(i)},y^{(i)}\}_{i=1}^N$ from the joint distribution of $(X,Y)$ \cite{Saharia2022Palette}, or alternatively by using a sample $\{x^{(i)}\}_{i=1}^N$ together with the likelihood $\L$ to design a so-called ``guidance'' term (see, e.g., \cite{Chung2022DPS,Kawar2022}).

\subsection{Denoising Diffusion Probabilistic Models}\label{sec:DDPM}
DMs can be formulated in various forms. For presentation clarity, we present the proposed MLMC approach by using Denoising Diffusion Probabilistic Models (DDPMs), a powerful class of DMs capable of generating approximate samples from arbitrary distributions. DDPMs have appeared in various forms and iterations since their first formulation in \cite{SohlDickstein2015}. DDPMs may be formulated directly as a finite sequence of noising kernels (each adding a discrete quantity of noise, hence here called discrete DDPMs) or as numerical approximations of a continuous time Stochastic Differential Equation (SDE, see Supplementary Material). For the purposes of illustration, we select the discrete formulation of e.g. \cite{Ho2020,Song2019,SohlDickstein2015,Song2020DDIM,Kawar2022} since it is extendable to non-Markovian models such as DDIM \cite{Song2020DDIM} and  it aligns more closely with the models used for our experiments in \cref{sec:Experiments}.
Recalling that our goal is to sample from the posterior, with density \(p(\cdot | y)\), for \(X\) given the observation \(Y=y\), we construct a sequence of ``states'' \(\lbrace{X_{i}}\rbrace_{i=0}^{T}\) with \(X_{0} = X\) and seek samples from the augmented posterior of \((X_{0}, \ldots, X_{T})\) given \(Y=y\), which admits the desired posterior as marginal. This joint posterior has density \(p_{0:T}(\ldots | y)\), which we decompose as
\begin{equation}
  p_{0:T}(x_0,\ldots,x_T|y)=\left(\prod_{t=1}^T \rKy_{t-1}(x_{t-1}\vert x_{t},y)\right) p_T(x_T|y)\,,\label{eq:DDPMdecomp}
\end{equation}
for given reverse transition kernels \((\rKy_{t})_{t=0}^{T-1}\)
and posterior density \(p_{T}(\cdot |y)\) of the final state \(X_{T}\).
%
%
%

Starting with the posterior of the final state, we capture several different constructions by setting \(p_{T}(x | y) = \phi_{n}(x ; (\mathbb I_{n} - \mathbb M) y, \mathbb M
)\)%
\footnote{We denote the standard normal density in $n$ dimensions by \(\phi_{n}(\cdot)\) and use the
  notation \(\phi_{n}(x ; \mu, \Sigma) = \phi_{n}( \Sigma^{-1/2} (x - \mu) )\) for \(\mu, x \in
  \R^{n}\), \(\Sigma \in \R^{n \times n}\), as the normal density with mean \(\mu\) and
  covariance matrix \(\Sigma\).} for some binary diagonal matrix \(\mathbb M\), which could potentially be degenerate. The choice of \(\mathbb M\) controls the dependence of the final state on the observation $y$.
We consider three cases that are frequently encountered in the literature:

\begin{enumerate}[align=parleft,left=0pt..2em]
\item \(\mathbb M = \mathbb I_{n}\), so that $p_{T}(\cdot \vert y) = \phi_{n}(\cdot)$, i.e.,
  \(X_{T}\) is a standard Gaussian independent of the observation, see, e.g., \cref{sec:SR3}. This choice
  can be motivated by relating \(X_{T}\) to the input \(X_{0}\) via
$$
p_{T}(x_T\vert y)= \frac{1}{ \mathcal E(y)} \int \K_{T;0}(x_T\vert x_0) \L (y\vert x_0,x_T) \pi(x_{0})  dx_0.
$$
where \(\mathcal E(\cdot)\) is the data evidence and \(\K_{T;0}\) is the forward transition
kernel from the input \(X_{0}\) to the final state \(X_{T}\). As we shall
shortly see, in DMs the transition kernel \(\K_{T;0}\) is chosen
to be a standard Gaussian independent of the value of the input, i.e.,
\(\K_{T;0}(\cdot | x_{0}) = \phi_{n}(\cdot)\). %
Then we choose the likelihood as $\L(\cdot \vert x_0,x_T)=\L(\cdot \vert x_0)$. That is, the
observation likelihood is independent of the final state given the input, to
yield $p_{T}(\cdot \vert y) = \phi_{n}(\cdot)$. \label{item:1}
\item \(\mathbb M=0\), i.e., we assume that for a given input, \(X_{T} = Y\) and hence
  we formally take $p_{T}(\cdot | y)=\delta_y(\cdot)$, see, e.g., \cref{sec:exp-denoising}. Here again the transition kernel
  \(\K_{T;0}(\cdot | x_{0})\) is standard Gaussian, hence this choice models the
  case when the conditional density of \(Y\) given \(X_{0}\) is standard
  Gaussian.
  \label{item:2}
\item If \(\mathbb M\) is a binary diagonal matrix with some ones and zeros on the
  diagonal, then the final state will depend on only parts of the observation, see, e.g.,  \cref{sec:exp-inpainting}.
  Typically, the same masking matrix, \(\mathbb M\), is used to define the transition
  kernels below, however we will omit this detail in the following for clarity
  of exposition.
  \label{item:3}
\end{enumerate}

Turning to the choice of transition kernels, one first considers a Gaussian,
Markovian forward process, for \(X_{t}\) given \(X_{t-1}\), whose transition
density is

\begin{equation}\label{eq:ForwardKernel}
  \K_{t}(x \vert x_{t-1}) = \phi_{n}\left(x; \sqrt{\frac{\gamma_t}{\gamma_{t-1}}} x_{t-1},\left(1-\frac{\gamma_t}{\gamma_{t-1}}\right)\mathbb{I}_{n}\right),
\end{equation}
for \(x \in \R^{n}\) and where $1=\gamma_0>\gamma_1>\ldots>\gamma_T>0$. %
This choice gives rise to a conditional distribution for \(X_{t}\) given the
input, \(X_{0}\), whose density is
\begin{equation}\label{eq:ForwardProcess}
    \K_{t;0}(x \vert x_{0}) = \phi_{n}(x; \sqrt{\gamma_t}x_0,(1-\gamma_t)\mathbb{I}_{n}),\quad 1\leq t\leq T,
\end{equation}
Note that $\K_{T;0}(\cdot\vert x_0) \to \phi_{n}(\cdot)$ when $\gamma_T \to 0$ as mentioned above.
Such Gaussian choices lead to tractable models for training and inference
\cite{Kawar2022,Song2020DDIM}.

We determine the reverse transition kernel by further
conditioning on the input, \(X_{0}\) and assuming that, given \(X_{0}\) and
\(X_{t}\), intermediate states are independent of the observation \(Y\), i.e.,
\begin{equation}\label{eq:reverse-kernel-cond-x0}
  \rKy_{t-1}(x | x_{t}, y) = \int
  \rK_{t-1;0,t}(x | x_{0}, x_{t})
  \rKy_{0;t} (x_{0} | x_{t}, y) d x_{0}.
\end{equation}
The reverse transition kernels $\rK_{t-1;0,t}(\cdot \vert x_0,x_t)=\phi_{n}(\cdot,
\mu_{t-1;0,t},\sigma_{t-1;0,t}^2\mathbb{I}_{n})$ where{, denoting $\Gamma_t=(1-\gamma_t)$,}
\cite{Ho2020}
\begin{equation}\label{eq:Sampling1}
\mu_{t-1;0,t}(x_0,x_t) = \sqrt{\frac{\gamma_t}{\gamma_{t-1}}}\frac{\Gamma_{t-1}}{\Gamma_{t}}(x_t-\sqrt{\gamma_t}x_0)+\sqrt{\gamma_{t-1}}x_0,\quad
\sigma_{t-1;0,t}^2=\frac{\Gamma_{t-1}}{\Gamma_{t}}\left(1-\frac{\gamma_{t}}{\gamma_{t-1}}\right).
\end{equation}
The only remaining unknown is the reverse transition kernel \(\rKy_{0;t} (\cdot |
x_{t}, y)\), for the input, \(X_{0}\), conditioned on the observation and
state \(X_{t}\). {Following \cite{Song2020}, we note that {the score
    function, \(\nabla\log \K_{t;0}(\cdot\vert x_0)\), satisfies} $\nabla\log \K_{t;0}(x\vert x_0)=
  (\sqrt{\gamma_t}x_0-x) / \Gamma_t$ and {use a neural network \(s_{\theta}\) with
    parameters \(\theta\) to approximate the score \(s_{\theta}(x, y, t ; \theta)\approx \nabla\log
    \K_{t;0}(x\vert x_0)\)}. Thus we obtain an estimate of \(x_{0}\) depending on
  the given state \(x_{t}\) and the observation, \(y\),
\begin{equation}\label{eq:x0param}
    x_0(x_t,y,t;\theta)=\frac{1}{\sqrt{\gamma_t}}(x_t+\Gamma_t\,s_{\theta}(x_t,y,t)),
  \end{equation}
  and replace the kernel \(\rKy_{0;t} (\cdot | x_{t}, y)\) with a point mass at
  that estimate, which when substituted back in
  \cref{eq:reverse-kernel-cond-x0} yields a Gaussian conditional, reverse
  transition kernel \(\rKy_{t-1}(x | x_{t}, y)\). Note that we introduce a
  dependence of \(s_{\theta}\) on $y$ since we want to approximate the reverse
  transition kernel $\rKy_{0;t}(\cdot \vert x_t,y)$, which in general depends on the
  observation. Alternative parametrisations, or indeed directly approximating
  $x_0$ are possible \cite{Ho2020,Song2020DDIM,Saharia2022}.} {As $s_{\theta}$
  approximates $\nabla\log K_{t;0}(x\vert x_0)$, (depending on the forward transition
  kernel), it is not suitable to use in reverse, i.e, starting from $X_T$,
  except in the specific case where the reverse process is also approximately
  Gaussian, which holds in the limit of a continuous process $\gamma_t\approx\gamma_{t-1}$
  \cite{SohlDickstein2015}. Therefore, producing accurate samples typically
  requires using a large number of intermediate steps. In particular, one
  generates} approximate samples \((x_{t})_{t=0}^{T-{1}}\) starting from
\(X_{T} \sim \mathcal N((\mathbb I_{n}- \mathbb M) y, \mathbb M)\), recursively
as follows
\begin{equation}\label{eq:DDPMUpdate}
  x_{t-1}=\sqrt{\frac{\gamma_{t-1}}{\gamma_t}}x_t-\sqrt{\frac{\gamma_{t}}{\gamma_{t-1}}}\left(\Gamma_{t-1}-\frac{\gamma_{t-1}}{\gamma_t}\Gamma_t\right)s_{\theta}(x_t,y,t)+\sqrt{\frac{\Gamma_{t-1}}{\Gamma_{t}}\left(1-\frac{\gamma_{t}}{\gamma_{t-1}}\right)}\xi_{t},
\end{equation}
where $\xi_{t}\sim\mathcal{N}(0,\mathbb I_{n})$. At the end, \(x_{0}\) will be an approximate sample from the
target posterior \(p(\cdot |
y)\). In practice, one incurs four sources of errors:
\begin{description}[style=unboxed,leftmargin=0.3cm]
\item[Truncation Error] Let \(p_{t}(\cdot | y)\) be the marginal density of
  \(X_{t}\) given \(Y=y\). The target posterior, $p(\cdot \vert y) = p_{0}(\cdot \vert y)$, is
  in general not supported on all of $\R^n$, whereas the Gaussian kernels
  $\rK_t(\cdot \vert x_t,y)$ and hence the densities, $p_t(\cdot \vert y)$, are. 
  In particular, the different supports of $p_0(\cdot\vert y)$ and $p_1(\cdot \vert y)$
  manifest in the fact that $s_{\theta}(x,t,y)$ will blow up as \(t \to 0\), 
  which causes numerical instability as \(t \to 0\) \cite{Song2020}. In practice, one either takes $x_1$ as an approximate sample from $p_0(\cdot |
  y)$ or uses a Gaussian approximation of the reverse transition kernel which
  does not use $s_{\theta}$, i.e., $\rK_0(\cdot \vert x_1) = \phi_{n}(\cdot \:;\:
  \sqrt{1/\gamma_1}x_1,\sigma_1^2)$, for a constant $\sigma_1 \geq 0$
  \cite{Song2020DDIM,Song2019}. In either case one incurs a `truncation error'. In other models, (e.g.
  \cite{Saharia2022}) altering the final step is not necessary, but one still implicitly assumes that a Gaussian posterior $p_0(\cdot | y)$ and
  thus incorporates the same error.

\item[Finite-Time Error] Note that for the forward process
  \cref{eq:ForwardProcess}, by choosing a sufficiently rapidly decreasing
  sequence with $\lim_{t \to T} \gamma_{t} = 0$, one obtains $\K_{T;0}(\cdot \vert x_0) \to
  \phi_{n}(\cdot)$, and sampling the final state, \(X_{T}\), can be done
  easily. %
  However, since one is going to reverse the process, one does not wish to
  have the difference between consecutive noise levels for the reverse process
  $\gamma_{t-1}/\gamma_{t}$ be too large, as this makes sampling the states too
  difficult, see \cref{eq:DDPMUpdate} and `Discretization Error' below. As a
  consequence, $\gamma_{t}/\gamma_{t-1}>0$ cannot be too small and so with a finite
  number of steps, one can only have $0<\gamma_T=\prod_{t=1}^T \gamma_{t}/\gamma_{t-1}$. In this
  case, sampling \(X_{T}\) as standard Gaussian induces an error, which we
  call a `finite-time' approximation error, as this is the same error that is
  incurred when diffusion models are obtained as finite-time approximations of
  the stationary solution of an SDE in continuous time.

\item[Model Error] 
  The parameters \(\theta\) of the mapping \(s_{\theta}\) are typically obtained by
  relying on the forward process \cref{eq:ForwardProcess} and a dataset of
  input/observation pairs $\left\{x_0^{(i)},y^{(i)}\right\}_{i=1}^N$, where
  the observations $y^{(i)}$ are given or generated from $x_0^{(i)}$ according
  to \cref{eq:InverseProblemNoisy0}. In this case, recalling
  \cref{eq:x0param}, one can obtain $\theta$ as
  \begin{gather}
    \arg\min_{\theta} \frac{1}{N\vert \mathcal{T}\vert}\sum_{t\in \mathcal{T}}\sum_{i=1}^N\left\lVert s_\theta(x_t^{(i)},t,y^{(i)})+\Gamma_t^{-1}(x_t^{(i)}-\sqrt{\gamma_t}x_0^{(i)})\right\rVert^2,\label{eq:LossFunctionDM}
  \end{gather}
  where $\mathcal{T}$ is a uniformly sampled set, with repetition, from \(\lbrace0, 1, \ldots,
  T\rbrace\).

  Here, when \(\mathbb M=\mathbb I_{n}\), \(x_t^{(i)}\) may be sampled given
  \(x_{0}^{(i)}\) via the transition kernel \(\K_{t;0}(\cdot |x_0^{(i)})\) in
  \cref{eq:ForwardProcess}, independently of the observation, \(y\); in
  particular \(x_{T}^{(i)}\) given \(x_{0}^{(i)}\) will be approximately
  standard Gaussian. Alternatively, when \(\mathbb M=0\), and following
  \cite{Liu2023}, one may instead sample \(x_{t}^{(i)}\) using the reverse
  transition kernel \(\rK_{t;0,T}(\cdot \vert x_0,x_T)=\phi_{n}(\cdot,
  \mu_{t;0,T},\sigma_{t;0,T}^2\mathbb{I}_{n})\)\footnote{One has, cf.
    \cref{eq:Sampling1}, \begin{equation*}
      \mu_{t;0,T}(x_0,x_T)= \sqrt{\frac{\gamma_T}{\gamma_t}}\frac{\Gamma_t}{\Gamma_T}(x_T -
      \sqrt{\gamma_{T}}x_{0})+ \sqrt{\gamma_t}
      x_{0} 
      ,\quad\sigma_{t;0,T}^{2}=\frac{\Gamma_t}{\Gamma_T}\left(1-\frac{\gamma_T}{\gamma_t}\right).
    \end{equation*}
    In the work \cite{Liu2023}, the authors use the so-called ``variance-exploding'' parametrisation, but following \cite[Appendix B]{Kawar2022}, their formulation may be shown to be equivalent to the above with the standard conversion between the variance-exploding parametrisation and the variance-preserving parametrisation used in this work.}
  and setting
  \(x_{T}^{(i)}=y^{(i)}\). %
%
  In this case, \(s_{\theta}\) would not need to directly depend on the observation
  \(y\), since for \(t=T\) we have \(x_{T} = y\) and for \(t<T\) the states
  \(x_{t}\) depend on \(x_{T}=y\).

  In either case, the error incurred by using the approximate $x_0$, as a
  function of $s_{\theta}$, in place of the true $x_0$ shall be called the `model
  error'.
\item[Discretization Error] Since one samples via \cref{eq:DDPMUpdate}, it is
  possible to `skip' steps and generate $x_{t-2}$ directly from $x_t$ reducing
  the NFEs by 1, or by \(T/2\) if every other step is skipped. %
  The reverse kernel $\rK_{t-2;t,0}( \cdot \vert x_{t},x_0)$ is derived via the decomposition
  \[
    \rK_{t-2;0, t}(x\vert x_0,x_t)=\int \rK_{t-1 ; 0,t}(x_{t-1}\vert
    x_0,x_t) \, \rK_{t-2;0,t-1}(x\vert x_0, x_{t-1}) \, dx_{t-1},
  \]
  which, using \cite[p.93]{Bishop}, yields $\rK_{t-2 ; 0, t}(\cdot \vert
  x_t,x_0)=\phi_{n}(\cdot ; \mu_{t-2 ; 0, t},\sigma_{t-2 ; 0, t}^2)$ and an update of the
  same form as \cref{eq:Sampling1}. In particular,
  \begin{equation*}
    \sigma_{t-2; 0, t}=\sigma^{2}_{t-2;0,t-1}+{\frac{\gamma_{t-1}}{\gamma_{t-2}}}\frac{\Gamma_{t-2}^2}{\Gamma_{t-1}^2}\sigma_{t-1;0,t}^2={\frac{\Gamma_{t-2}}{\Gamma_{t}}\left(1-\frac{\gamma_{t}}{\gamma_{t-2}}\right)}.
  \end{equation*}
  %
  Similarly, for any integer \(M>1\),
  \begin{equation}\label{eq:variancecoarse}
    \sigma_{t-M ; 0, t}^2=\sigma^{2}_{t-M;0,t-M+1}+{\frac{\gamma_{t-M+1}}{\gamma_{t-M}}}\frac{\Gamma_{t-M}^2}{\Gamma_{t-M+1}^2}\sigma_{t-M+1;0,t}^2={\frac{\Gamma_{t-M}}{\Gamma_{t}}\left(1-\frac{\gamma_{t}}{\gamma_{t-M}}\right)},
  \end{equation}
  where the last equality follows by induction.
  Sampling would be implemented iteratively via an update of the same
  form as \cref{eq:DDPMUpdate}, with a reduced number of steps,
  \begin{equation}\label{eq:coarsesampling}      x_{t-M}=\sqrt{\frac{\gamma_{t-M}}{\gamma_{t}}}x_t-\sqrt{\frac{\gamma_{t}}{\gamma_{t-M}}}\left(\Gamma_{t-M}-\frac{\gamma_{t-M}}{\gamma_{t}}\Gamma_t\right)s_{\theta}(x_t,t,y)+\sqrt{\frac{\Gamma_{t-M}}{\Gamma_{t}}\left(1-\frac{\gamma_{t}}{\gamma_{t-M}}\right)}\xi_{t},
  \end{equation}
  where $\xi_{t}\sim\mathcal{N}(0,\mathbb{I}_n)$. While skipping steps reduces the overall NFEs
  and hence the overall cost of sampling, it also introduces the error which
  we call the discretization error; motivated again by the induced error when
  diffusion models are obtained via a time discretisation of an SDE in
  continuous time. To see the source of this error in our discrete setting,
  recall that we approximate the reverse transition kernel \(\rKy_{0;t}(\cdot, |
  x_{t}, y)\) in \cref{eq:reverse-kernel-cond-x0} by a point mass function at
  an estimate of \(x_{0}\), which leads to a Gaussian approximation of the
  reverse transition kernel, \(\rKy_{t-1}(x | x_{t}, y)\). Hence, we expect
  that when skipping steps, the reverse transition kernel \(\rKy_{t-M}(x |
  x_{t}, y)\) will be further away from a Gaussian compared to \(\rKy_{t-1}(x
  | x_{t}, y)\) \cite{SohlDickstein2015}.
\end{description}
For our purposes then, the neural network approximating the score and the
truncation parameter error are taken to be fixed; the effect of finite-time
error is negligible in practice compared to the other error
sources\footnote{{ In other words, we will be using the ``inexact'' model
    with these fixed parameters as the ``true'' model. One could use MLMC to
    reduce the finite-time error via a hierarchy of approximations with
    different finite times. However, since a relatively small finite time is
    already sufficient to obtain near-accurate results, such an approach would
    not significantly reduce the total computational cost.}}
\cite{Benton2023}. Thus the most significant error, and the one we are
concerned with in this work, is the discretisation error. We note that
modifications to the noise or score function scaling in \cref{eq:DDPMUpdate},
such as DDIM \cite{Song2020DDIM}, and concatenating together several steps as
described above, have been used to accelerate sample generation (whilst still
training on a finer, possibly continuous time mesh). As has been emphasised,
this is because, while \(10^{3}\) steps produce accurate samples, generating
them is too slow for most Bayesian computation applications. The power of the
proposed MLMC-based acceleration is that it combines the rapidity of cheap
samples with the precision of expensive ones when computing Monte Carlo
averages, {since, due to the telescoping structure of MLMC, errors that
  are made at coarser discretisation levels are cancelled when a new level is
  introduced.}

\def\aX{\widehat X}
\subsection{Multilevel Monte Carlo}\label{sec:MLMC}
Conventional Monte Carlo averaging delivers estimates
of (analytically intractable) expectations of some function $f$ of a random
variable \(X\), with probability density \(p(\cdot)\), via the sum
\begin{equation}\label{eq:MC}
  \mathbb{E}[f(X)]=\int f(x)p(x) dx \approx N^{-1}\sum_{i=1}^N f(x^{(i)}),
\end{equation}
where $x^{(i)}$ are independent realisation of \(X\) and where
the integral is over the domain of \(X\). The variance of the estimate is then
$N^{-1}\mathbb{V}[f(X)]=N^{-1}\mathbb{E}\left[\lVert
  f(X)-\mathbb{E}[f(X)]\rVert^2\right]$. As a consequence, obtaining
an estimate with {root mean-square error} $\varepsilon$ requires $N=\mathcal{O}(\varepsilon^{-2})$ samples,
which may be prohibitive if sample generation is costly. This is precisely
the case with DMs for Bayesian inversion in computational imaging, where one may require $10^3$ NFEs per sample $x^{(i)}$
to obtain accurate inferences
\cite{Song2020}. However, whilst accurate samples may require $10^3$ NFEs or more, in many cases one can obtain meaningful inferences with as little as 50 NFEs (see, e.g., \cref{fig:sr3samples}). MLMC \cite{Giles2015} seeks
to exploit this by appropriately combining samples with coarse resolution with
samples with fine resolution, while maintaining the overall accuracy of the
estimate, at a fraction of the cost. More concretely, consider approximations
of \(X\), denoted by $\lbrace \aX_{\ell} \rbrace_{\ell=0,1\ldots,L}$\footnote{We note the slight
  abuse of notation. In the previous section, \(\lbrace X_{i}\rbrace_{i=0}^{T}\) denoted
  the states of the diffusion model, starting with \(X_{0}=X\). In the current
  section, \(\lbrace \aX_{\ell}\rbrace_{\ell=0}^{L}\) denote approximations of \(X\). }, where
the resolution and the cost of generating a sample of \(\aX_{\ell}\), e.g., the
number of NFEs, increase with \(\ell\). For example, take that a sample of
$\aX_{L}$ requires $10^3$ NFEs, while a sample of $\aX_{0}$ requires $50$
NFEs, with intermediate values of $\ell$ requiring an increasing sequence of NFEs
per sample.
%

We seek to approximate the expectation \(\mathbb{E}[f(X)]\). One could do this
in a similar fashion as in \cref{eq:MC} only using samples $\widehat x_{L}^{(i)}$ of
\(\aX_{L}\) for sufficiently large \(L\). Alternatively, one may build an
estimator, $Y$, for $\mathbb{E}[f(\aX_L)]$ using the identity
\begin{align}
  \mathbb{E}[f(\aX_L)]&=\mathbb{E}[f(\aX_0)]+\sum_{\ell=1}^L\mathbb{E}[f(\aX_{\ell})-f(\aX_{\ell-1})]\nonumber\\
    &\approx N_{0}^{-1}\sum_{i=1}^{N_{\ell}}f(\widehat x_0^{(0,i)})+\sum_{\ell=1}^LN_{\ell}^{-1}\sum_{i=1}^{N_{\ell}}\left[f(\widehat x_{\ell}^{(\ell,i)})-f(\widehat x_{\ell-1}^{(\ell,i)})\right]\nonumber\\
                    &\equiv Y_{0} + \sum_{\ell=1}^LY_{\ell}\equiv Y \label{eq:MLMCEstimator}.
\end{align}
Here, \(\lbrace {\widehat x_{\ell}^{(\ell,i)}, {\widehat x_{\ell-1}^{(\ell,i)}}} \rbrace_{\ell, i} \) are correlated
samples of \(\lbrace {\aX_{\ell}, \aX_{\ell-1}} \rbrace\). This correlation is crucial so that the
variance of the difference \(f(\aX_{\ell}) - f(\aX_{\ell-1})\) gets smaller, hence
requiring fewer samples, as \(\ell\) increases and the cost of sampling
this difference increases. Denoting the cost per sample and variance of
\(f(\aX_{\ell}) - f(\aX_{\ell-1})\) by \(C_{\ell}\) and $V_{\ell}$ respectively, and
$C_{0},V_0$ the cost and variance of $f(\aX_0)$, the total cost is
$C_{\textnormal{MLMC}}=\sum_{\ell=0}^LN_{\ell}C_{\ell}$ and the total variance is
$\mathbb{V}[Y]=\sum_{\ell=0}^LN_{\ell}^{-1}V_{\ell}$. Minimizing the total cost subject to \(V[Y ]<\varepsilon^{2}/2\) yields the optimal number of
samples
\begin{equation}\label{eq:algo_Nl}
  N_{\ell} \approx 2 \varepsilon^{-2}\sqrt{V_{\ell}C_{\ell}^{-1}}
  \left(\sum_{\ell=0}^L\sqrt{V_{\ell}C_{\ell}}\right),
\end{equation}
and the optimal total cost $C \approx
\varepsilon^{-2}(\sum_{\ell=0}^L\sqrt{V_{\ell}C_{\ell}})^2$ \cite{Giles2015}. Hence the total cost
is proportional to
\[
  \begin{cases}
    \varepsilon^{-2}\, V_{0}\, C_{0}  & \textnormal{ if }  V_{\ell} C_{\ell} \textnormal{ decreases sufficiently fast } \\
    \varepsilon^{-2}\, L^{2} & \textnormal{ if } V_{\ell} C_{\ell} \approx \textnormal{constant} \\
    \varepsilon^{-2}\, V_{L}\, C_{L} & \textnormal{ if } V_{\ell} C_{\ell} \textnormal{ increases sufficiently fast.}
  \end{cases}
\]
For example, compare the first case to the total cost of standard Monte Carlo
sampling of \(\mathbb{E}[f(\aX_L)]\) which is \( \propto \varepsilon^{-2}\,
\mathbb{V}[f(\aX_{L})] \, C_{L}\). Since \(C_{0} \ll C_{L}\) and typically
\(V_{0} \approx \mathbb{V}[f(\aX_{L})]\) --- as we do not expect a significant
dependence of variance of the quantity on the approximation if the used random
variables are appropriately scaled --- the cost of MLMC is significantly
smaller.


In sample generation using DMs, the sample cost and accuracy
increases as the number of time steps, $T$, in \cref{eq:DDPMdecomp} increases,
which is equivalent to the NFEs. We set the number of time steps for
\(\aX_{\ell}\) to be $T_0M^{\ell}$ for some integers \(T_{0}, M > 1\), the cost for
generating a sample of \(f(\aX_{\ell}) - f(\aX_{\ell-1})\) at each level \(\ell\) is
then $C_{\ell}=C_0(M^{\ell}+M^{\ell-1})=C_0(1+M^{-1})M^{\ell}<2C_0M^{\ell}$, for a constant
\(C_0\). In practice, we additionally have, or assume, that \(V_{\ell} \propto M^{-\beta
  \ell}\) and \( \lVert \mathbb{E} [{f(X) - f(\aX_{\ell})} \rVert] \propto M^{-\alpha \ell}\) for
some \(\alpha, \beta > 0\). Given that the MSE is
\begin{equation}\label{eq:MSE}
  \textnormal{MSE} = \mathbb{E}[\lVert Y-\mathbb{E}[f(X)]\rVert^2] = \mathbb{V}[Y] + \lVert\mathbb{E}[f(\aX_{L})-f(X)]\rVert^2,
\end{equation}
we impose \(\lVert \mathbb{E}[f(\aX_{L})-f(X)] \rVert^{2} \leq \varepsilon^{2}/2\) yielding
$L \approx \frac{-1}{\alpha}\lvert \log_M (\varepsilon/2)\rvert$, up to a constant. Substituting
the value of \(L\) and the bounds of \(\lbrace C_{\ell}, V_{\ell} \rbrace_{\ell=0}^{L}\) in the
total costs, we arrive at \cite{Giles2015} 
\begin{equation}\label{eq:MLMCAsymptoticCost}
  C_{\textnormal{MLMC}} \lesssim
  V_{0} \, C_{0} \, \varepsilon^{-2}
  \begin{cases}
    1 & 1<\beta,\\
    \lvert \log_M(\varepsilon) \rvert^2 & 1=\beta,\\
    \varepsilon^{(\beta - 1) /\alpha}, & 1>\beta.
  \end{cases}
\end{equation}
Hence, comparing the cost of classical Monte Carlo estimators of
\(\mathbb E [f(X)]\) for the same accuracy, which is of order
\(\varepsilon^{-2-1/\alpha}\), MLMC offers significant savings with the best case being when
\(\beta > 1\).

\paragraph{An algorithm for MLMC} A heuristic algorithm which iteratively
increments $\lbrace N_{\ell} \rbrace_{\ell=0}^{L}$ and $L$ to reduce the bias and variance below
a given tolerance is shown in \cref{alg:MLMC}, see also \cite{Giles2015}. The
bias estimate may be taken as $\lVert\mathbb{E}[Y-f(\aX_{L})]\rVert \approx \lVert
Y_{L}\rVert / (M^\alpha-1)$, see \cite{Giles2015}.
\begin{remark}[Strategies for estimating the bias]
  When estimating the bias, we typically utilise $Y_{L-1}$ as well as $Y_L$,
  as $N_{\ell}$ decreases with $\ell$, and hence the calculation of $Y_{L}$ may use
  fewer samples and be less stable than the calculation of \(Y_{L-1}\). For
  example, we may check that $\max\left(\lVert Y_{L-1}\rVert M^{-\alpha},\lVert
    Y_{L}\rVert\right) / (M^\alpha-1)$. Similarly, since one has access to $\alpha,\beta$
  (via linear regression of $V_{\ell}$ and \(\lVert Y_{\ell}\rVert\) or theoretical
  analysis), one may use the models $V_{\ell}\propto M^{-\beta\ell}, \lVert Y_{\ell}\rVert \propto
  M^{-\alpha\ell}$ to prevent underestimation of the variance and bias
  \cite{Giles2015}. Finally, noting that the norm $\lVert Y_{\ell}\rVert$ is a
  biased estimator of $\|\mathbb{E}[f(\aX_\ell)-f(\aX_{\ell-1})]\|$, we have found
  that debiasing the estimator produces better results when the number of
  samples is small. We do this by iteratively scaling the biased estimate
  $\lVert Y_{\ell}\rVert$ by the factor $\left(1+N_{\ell}^{-1}V_{\ell}/\lVert
    Y_{\ell}\rVert^2\right)^{1/2}$ and recomputing $V_{\ell}$ (which depends on
  $Y_{\ell}$) until the estimate converges.
\end{remark}

\paragraph{Choice of starting level, $\ell_{0}$} The starting value $\ell_{0}$ one uses can be determined via Eq. (3.6) in \cite{Giles2019}
\begin{equation}\label{eq:Optimal_l0}
    V_{1}\leq\frac{\left(\sqrt{\mathbb{V}[f(\aX_{1})]M}-\sqrt{\mathbb{V}[f(\aX_{0})]}\right)^2}{(1+M)},
\end{equation}
where one generates samples to estimate $V_{1}=\mathbb{V}[f(\aX_{1})-f(\aX_{0})]$
and $\mathbb{V}[f(\aX_1)]$ with $M^{\ell_0+1}$ steps, and $\mathbb{V}[f(\aX_0)]$ with
$M^{\ell_0}$ steps. To maximise the efficiency of MLMC one increases $\ell_{0}$
until one fulfills \cref{eq:Optimal_l0}.

\subsection{Sample Generation with Diffusion Models for MLMC}
In our setting, the MLMC samples which are used in \cref{alg:MLMC} are
generated via DMs {using \cref{eq:coarsesampling} iteratively to give an
  correlated samples of $(\aX_{\ell}, \aX_{\ell-1})$.
  \cref{alg:ContinuousCoarseFine} details a single iteration to generate such
  samples. Given an approximation level, \(\ell\), and starting from equal coarse
  and fine states \(\widehat x_{\ell-1}=\widehat x_{\ell}\), sampled from a normal
  distribution, cf.~\cref{sec:DDPM}, one updates this pair of states by
  iteratively calling \cref{alg:ContinuousCoarseFine} a total of \(M^{\ell-1}\)
  times.}

For MLMC to have better computational efficiency than Monte Carlo, the
correlation at each level $\ell$ between the samples $\aX_{\ell}$ and $\aX_{\ell-1}$
should be maximised. To accomplish this, one uses the same realisation of the
Gaussian noise increments $(\xi_{t})_{t=0}^{T}$ with two different
discretisations to generate each sample \cite{Giles2015}. In practice this is
accomplished by summing the same set of Gaussian noise increments $\xi_{t}$ for
the fine and coarse paths. However, this summing must be done in a weighted
manner since from \cref{eq:variancecoarse} it is clear that the variance of
the added noise for a coarse step (assuming it is equivalent to two fine
steps) is a weighted sum of the variances for the two fine steps - thus one
may generate the coarse noise $\eta^{(t)}_c$ via two fine noise samples
$\eta^{(t)}_f=\sigma_{t-1;0,t}\xi_t$ using $
\eta^{(t)}_c=\sqrt{\frac{\gamma_{t-1}}{\gamma_{t-2}}}\frac{\Gamma_{t-2}}{\Gamma_{t-1}}\eta^{(t)}_f+\eta^{(t-1)}_f.
$ Then, from \cref{eq:variancecoarse} in the general case of $M$ fine steps to
every coarse step, one updates $\eta^{(t)}_c$ according to
\cref{line:noisereweighting} in \cref{alg:ContinuousCoarseFine}. $\eta^{(t)}_c$
is then maximally correlated with the fine path and has the correct scale.
This is then used the reverse the coarse state in \cref{line:coarsestep}.

\begin{algorithm}
\caption{MLMC Algorithm}
\label{alg:MLMC}
    \textbf{Input}: $\ell_0,N^{(0)}$
  \begin{algorithmic}[1]
  \State $L\gets \ell_0+2$
\State Set target samples to $N^{(0)}$ for $\ell_0 \ldots L$
\While{extra samples required}
\State Generate extra samples at each level
 \State Estimate $V_{\ell}, \lVert Y_\ell\rVert$ for \(\ell = \ell_0 \ldots L\)
  \State Update required number of samples $N_{\ell}$ for \(\ell = \ell_0 \ldots L\) using \cref{eq:algo_Nl}
\If{bias larger than $\varepsilon/\sqrt{2}$}
 \State$L\xleftarrow{}L+1$
  \State Estimate $V_{L}$ and calculate $N_L$
\EndIf
\EndWhile
\end{algorithmic}
\end{algorithm}

\begin{algorithm}
    \caption{Reverse fine and coarse states}\label{alg:ContinuousCoarseFine}
    \textbf{Input}: $\widehat x_{\ell},\widehat x_{\ell-1},t ,M,\ell,L$

    \emph{Assume the two states, \(\widehat x_{\ell}\) and \(\widehat x_{\ell-1}\),
      are at time \(t\). Reverse \(\widehat x_{\ell}\) by \(M\) steps, each of
      size \(M^{L-\ell}\), and reverse \(\widehat x_{\ell-1}\) by one step of
      size \(M^{L-\ell+1}\).}
    \begin{algorithmic}[1]
      \State $\eta_{c}\gets 0$
      \State $\Delta_\ell \gets M^{L-\ell}$
      \For{$i=1,\ldots, M$}
        \State $\eta_f\sim\mathcal{N}(0,\sigma_{t-\Delta_\ell ; 0, t})$
        \State $\widehat x_{\ell}\gets\sqrt{\frac{\gamma_{t-\Delta_{\ell}}}{\gamma_{t}}}\widehat x_{\ell}-\sqrt{\frac{\gamma_{t}}{\gamma_{t-\Delta_{\ell}}}}\left(\Gamma_{t-\Delta_{\ell}}-\frac{\gamma_{t-\Delta_{\ell}}}{\gamma_{t}}\Gamma_{t}\right)s_{\theta}(\widehat x_{\ell},t,y)+\eta_f$,{\hfill see \cref{eq:coarsesampling}}
        \State $\eta_c\gets \sqrt{\frac{\gamma_{t}}{\gamma_{t-\Delta_{\ell}}}}\frac{\Gamma_{t-\Delta_{\ell}}}{\Gamma_{t}}\eta_c+\eta_f$\label{line:noisereweighting}
        \State $t\gets t-\Delta_\ell$
      \EndFor
      \State $\Delta_{\ell-1} \gets M \Delta_{\ell}$
      \State \(t \gets t + \Delta_{\ell-1}\)
      \State $\widehat x_{\ell-1}\gets\sqrt{\frac{\gamma_{t- \Delta_{\ell-1}}}{\gamma_{t}}}\widehat x_{\ell-1}-\sqrt{\frac{\gamma_{t}}{\gamma_{t-\Delta_{\ell-1}}}}\left(\Gamma_{t-\Delta_{\ell-1}}-\frac{\gamma_{t-\Delta_{\ell-1}}}{\gamma_{t}}\Gamma_{t}\right)s_{\theta}(\widehat x_{\ell-1},t,y) +\eta_c$,{\hfill see \cref{eq:coarsesampling}}\label{line:coarsestep}
      \State \Return $\widehat x_{\ell},\widehat x_{\ell-1}$
    \end{algorithmic}
\end{algorithm}

\section{Experiments}\label{sec:Experiments}
We now demonstrate the effectiveness of the proposed MLMC approach with three canonical inverse problems related to computational imaging: image denoising, image inpainting, and image super-resolution. We have chosen these specific problems because they allow us to demonstrate the MLMC approach with three markedly different DM-based posterior sampling strategies that are representative of the main strategies studied in the literature\footnote{The code required to reproduce our experiments is
available from \url{https://github.com/lshaw8317/MLMCforDMs}.}.

{Since the quantitative and scientific imaging applications of interest rely strongly on the posterior mean estimator (i.e. minimum mean square error solution), the posterior covariance is a natural measure for uncertainty quantification. Unfortunately, computing the full posterior covariance is infeasible in imaging problems because of the high dimensionality involved. Therefore, to probe the posterior covariance, one typically performs a multiresolution analysis, first computing the marginal posterior variance of individual
image pixels, and subsequently of groups of pixels of size $2\times 2,4\times 4,\ldots,32\times 32$ pixels \cite{Laumont2022}. One thus captures the uncertainty in image structures of different sizes and in different regions of the scene, and in a manner that is easy to visualize. Of all these uncertainty visualization maps, the map corresponding to the individual pixel scales is the most expensive to compute, as individual pixels
exhibit higher variance than groups of multiple pixels. Hence, we choose to use the marginal second moment of the individual image pixels to measure the gain in computational efficiency of MLMC relative to the conventional MC approach.} The
quantity of interest is then simply $f(x)= x^{2}$ for \(x \in \R^{n}\), where
the square of a vector is defined element-wise.

We compare MLMC to MC via the ratio of the MSEs when approximating the second
moment for the same cost, defined as total number of NFEs.
%
In all experiments we select the refinement factor between levels as $M=2$.
Recall that the maximum level required, $L\approx-\log_M(\varepsilon)/\alpha$, is such that the
bias be less than $\varepsilon/\sqrt{2}$. We therefore expect that, for a given $\varepsilon$, the
efficiency gain of MLMC relative to MC to be greater the smaller $\alpha$. To
obtain estimates of the coefficients $\alpha$ and $\beta$ we use estimates of $\lVert
Y_{\ell}\rVert$ and $V_{\ell}$, respectively. {These are also used to determine the starting value $\ell_0$ via \cref{eq:Optimal_l0}.} {For our numerical study of MLMC, we use either $10^3$ (image super-resolution and inpainting experiments) or $10^4$ (image denoising experiment)  samples at each level to obtain estimates of the sample variance $\lbrace V_\ell \rbrace_{\ell}$ and bias $\lbrace Y_\ell \rbrace_{\ell}$. From these estimates we obtain estimates of $\alpha$ and $\beta$ by linear regression.} Using these estimates, we
verify our MLMC methods by checking whether the desired accuracy is indeed obtained, via
\begin{equation}\label{eq:eps_est}
\varepsilon_{\text{est}}=\sqrt{\lVert Y_{L}\rVert^2/(M^{\alpha}-1)^2+\sum_{\ell=0}^LV_{\ell}/N_{\ell}}.
\end{equation}

\begin{figure}
\centering
    \begin{subfigure}[t]{.1\textwidth}
        \includegraphics[width=\textwidth]{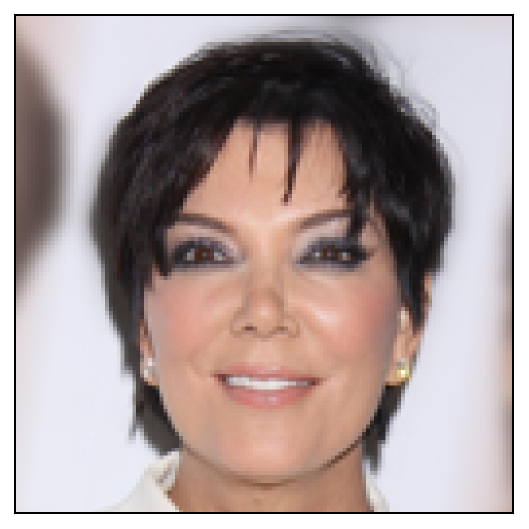}
    \end{subfigure}
    \begin{subfigure}[t]{.1\textwidth}
        \includegraphics[width=\textwidth]{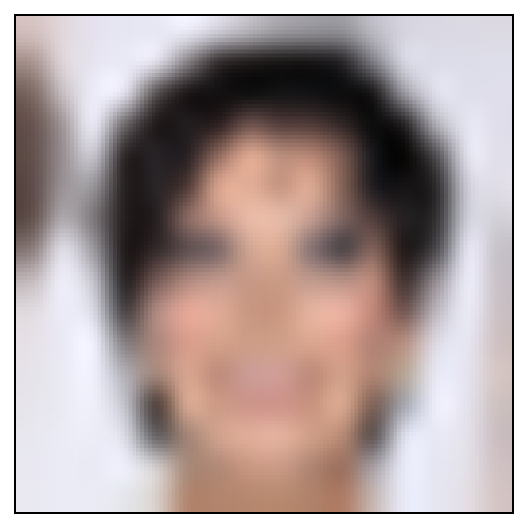}
        \caption{SR}
        \label{fig:sr}
    \end{subfigure}
    \begin{subfigure}[t]{.1\textwidth}
        \includegraphics[width=\textwidth]{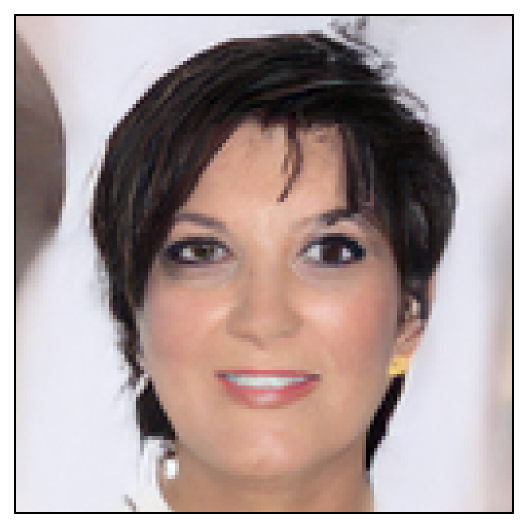}
    \end{subfigure}
        \begin{subfigure}[t]{.1\textwidth}
        \includegraphics[width=\textwidth]{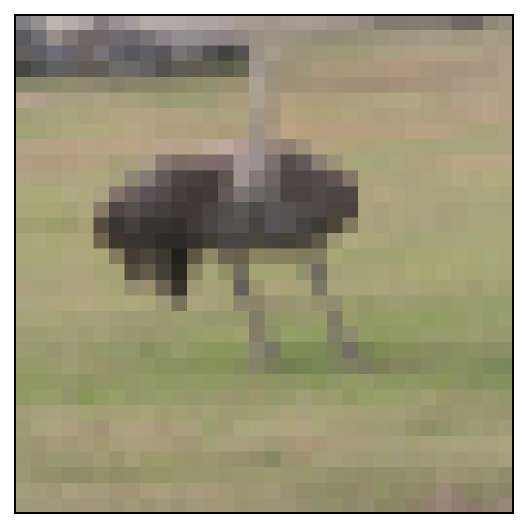}
    \end{subfigure}
    \begin{subfigure}[t]{.1\textwidth}
        \includegraphics[width=\textwidth]{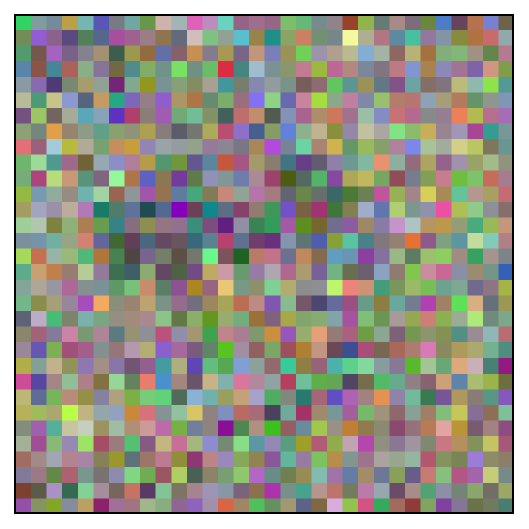}
        \caption{DN}
        \label{fig:dn}
    \end{subfigure}
    \begin{subfigure}[t]{.1\textwidth}
        \includegraphics[width=\textwidth]{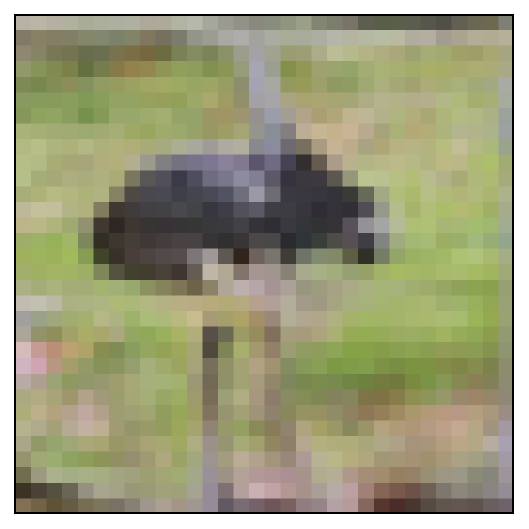}
    \end{subfigure}
        \begin{subfigure}[t]{.1\textwidth}
        \includegraphics[width=\textwidth]{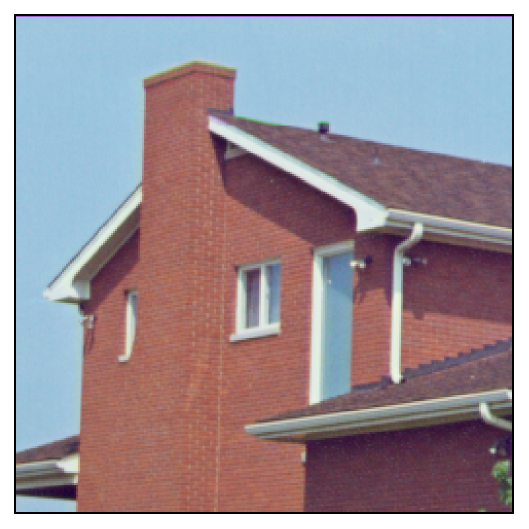}
    \end{subfigure}
    \begin{subfigure}[t]{.1\textwidth}
        \includegraphics[width=\textwidth]{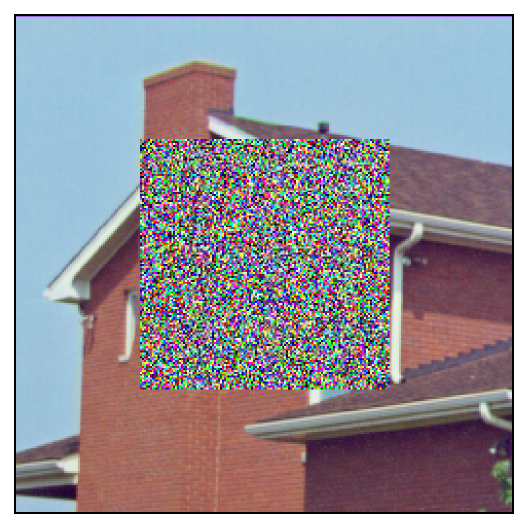}
        \caption{IP}
        \label{fig:ip}
    \end{subfigure}
    \begin{subfigure}[t]{.1\textwidth}
        \includegraphics[width=\textwidth]{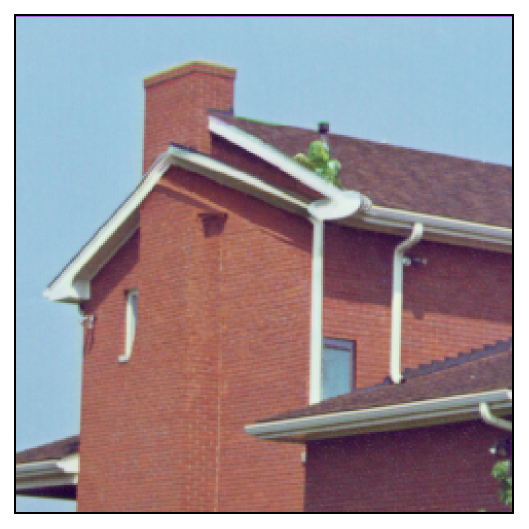}
    \end{subfigure}
    \caption{Three imaging inverse problems: super-resolution (SR), denoising (DN),  and inpainting (IP). For each we show, left to right: truth $x$; observation $y$; posterior sample from a DM.}
        \label{fig:invprobs}
\end{figure}

\subsection{Experiment 1: Image super-resolution}
\label{sec:SR3}
We first demonstrate the effectiveness of our proposed approach in a Bayesian
image super-resolution problem. In a manner akin to \cite{Saharia2022}, we
consider \cref{eq:InverseProblemNoisy} in the noiseless setting, $\eta=0$, and
thus $y=\mathcal{A}x$, \(\mathcal{A} \in \R^{m\times n}\). We focus on the $16\times16\to128\times128$ case, where
$y$ is an image of size $16\times16$ pixels obtained by down-sampling a
high-resolution image $x$ of size $128\times128$ pixels (see \cref{fig:sr}). This
is a challenging ill-posed inverse problem, as the dimension of $y$ is
dramatically smaller than the dimension of $x$, leading to significance
posterior uncertainty. \cref{fig:sr} depicts a sample from the posterior
distribution $p(\cdot\vert y)$, as generated by the machine-learning-based SR3
model\footnote{ We use the implementation
  \url{https://github.com/Janspiry/Image-Super-Resolution-via-Iterative-Refinement}
  of SR3. Note we use the SR3 model, not the DDPM model, from the given
  codebase. This implementation modifies the noise scheduling
  that appears in the paper; all schemes agree in the
  limit $\frac{\gamma_{t}}{\gamma_{t-1}}\to1$.} with 2048 steps \cite{Saharia2022}. The
SR3 model is a conditional DM obtained by training a network $s_{\theta}(x,t,y)$
specialised for image superresolution, with the loss function
\cref{eq:LossFunctionDM} and a dataset of paired high ($x_0$) and low ($y$)
resolution images $\{x_0^{(i)},y^{(i)}\}_{i=1}^N$. Note that in this case one
has $\mathbb{M}=0$ and so $p_T(x\vert y)=\phi_n(x;0,\mathbb{I}_n)$ (see \cref{item:1}
in \cref{sec:DDPM}). Therefore, for sampling from the posterior $p(\cdot\vert y)$
one proceeds to generate samples via \cref{eq:DDPMUpdate} with
$x_T\sim\mathcal{N}(0,\mathbb{I}_n)$.

\begin{figure}
\centering
    \includegraphics[width=\textwidth]{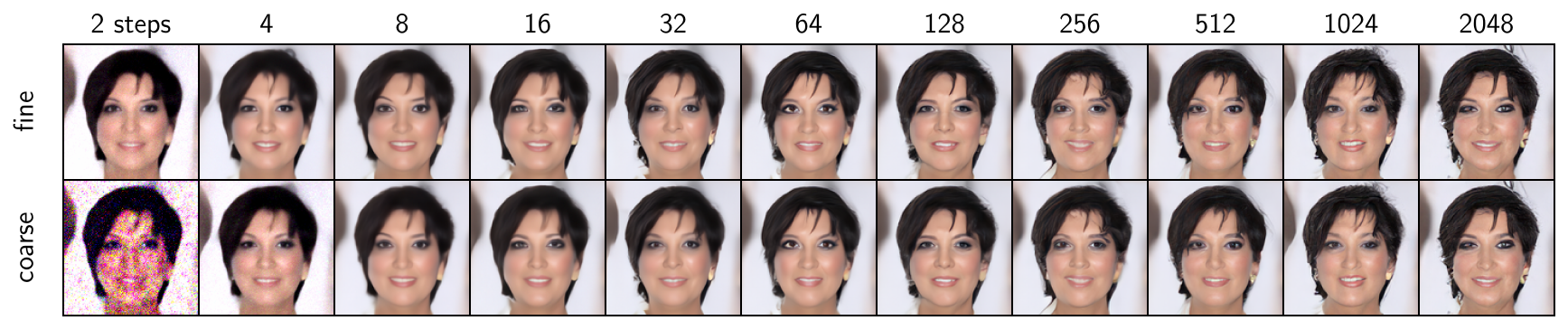}
    \caption{Sample images for the super-resolution problem at the coarse and fine levels. Since the coarse and fine path start from the same image and are driven by the same random noise path, the samples in each column are well correlated, leading to improved MLMC efficiency.}
    \label{fig:sr3samples}
\end{figure}

To illustrate the computational efficiency of MLMC relative to the conventional MC approach, we use MLMC to estimate the marginal second moment of each image pixel to various degrees of precision. Results are shown in \cref{fig:Superresolution}, where one sees that MLMC substantially outperforms MC (e.g., to obtain a precision of 0.0013, corresponding to $0.1\%$ of the pixel dynamic range, it is $4\times$ more efficient to use MLMC rather than MC, where one uses $2^{10}=512$ NFEs at the finest level). The variance in the upper left plot is especially large for small $\ell$ since the sampler starts from $\mathcal{N}(0,I)$; hence with a few steps, this noise is not removed and thus the variance over a set of samples is large. Correspondingly, correlation between the fine and coarse paths is null and so the variance of the difference $f(\aX_{\ell})-f(\aX_{\ell-1})$ is essentially $\mathbb{V}[f(\aX_{\ell})]+\mathbb{V}[f(\aX_{\ell-1})]$. As a consequence, the condition
\cref{eq:Optimal_l0} is not satisfied for a number of steps less than 32. Turning to the upper-right plot, one has $\alpha=0.7$. \cref{fig:sr3samples} shows example samples at various levels.

\subsection{Experiment 2: Image denoising}\label{sec:exp-denoising}
We now study the Gaussian image denoising problem $y=x_0+\sigma_y\eta$ with
$\eta\sim\mathcal{N}(0,\mathbb{I}_n)$, resulting from \cref{eq:InverseProblemNoisy} with $\mathcal{A} =
\mathbb I_{n}$. We consider a high noise level $\sigma_y=0.8$ (image pixels take values
in the range $[-1,1]$), so there is significant posterior uncertainty as a
result. DM-based Bayesian methods for this problem leverage the fact that $y$
is a noisy version of $x_0$, so one may scale $y$ such that the noise level
corresponds to that of the solution at time $T$ of the forward noising process
\cref{eq:ForwardProcess} as detailed in \cite[Appendix
B]{Kawar2022}. 
We thus obtain
$$
{y}^\dagger=\sqrt{\gamma_{T}}y=\sqrt{\gamma_{T}}x_0+\sqrt{\gamma_{T}}\sigma_y\eta,
$$
so
$
\gamma_{T}=(1+\sigma_y^2)^{-1}.
$
To perform Bayesian computation, one may generate approximate samples from the posterior distribution $p(\cdot\vert y)$ via the reverse process \cref{eq:DDPMUpdate} with $p_T(\cdot \vert y)=\delta_{{y}^\dagger}(\cdot)$\footnote{The same noising process is applied by the forward kernel and the noise associated with the inverse problem, so that $y=x_T$. Note it also coincides with the choice $\eta=1$ in eq. (7) in \cite{Kawar2022}.}. In this case, the mapping $s_{\theta}$ is independent of $y$, since $\rKy_{t-1}(\cdot\vert x_t,y)$ is independent of $y$ ($y$ is always noisier than $x_t$ for $t<T$). In our Bayesian imaging denoising experiment, $s_{\theta}$ is a neural network specialised for image denoising, trained on the CIFAR10 data by using a VPSDE formulation\footnote{We use the implementation \url{https://github.com/yang-song/score_sde_pytorch}. Note that \cite{Song2020} uses the SDE framework for DMs - the following presentation adapts the formulation used there to the discrete form.}. \cref{fig:dn} presents an example of a clean CIFAR10 image and its noisy observation, \(y\), with $\sigma_y=0.8$, together with a posterior sample as generated by the DM of \cite{Song2020} with 2048 steps.

\begin{figure}
\centering
\begin{subfigure}{0.8\textwidth}
    \includegraphics[width=\textwidth]{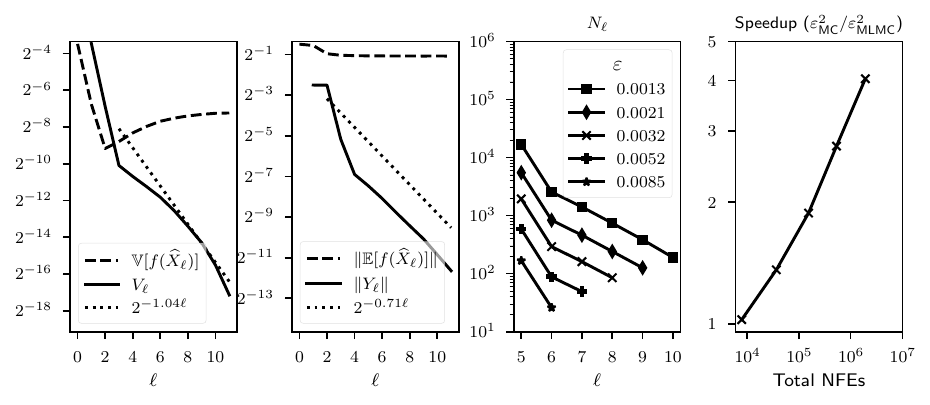}
    \caption{Results for the superresolution (SR) experiment in \cref{sec:SR3}}
    \label{fig:Superresolution}
    \end{subfigure}
    \begin{subfigure}{0.8\textwidth}
    \includegraphics[width=\textwidth]{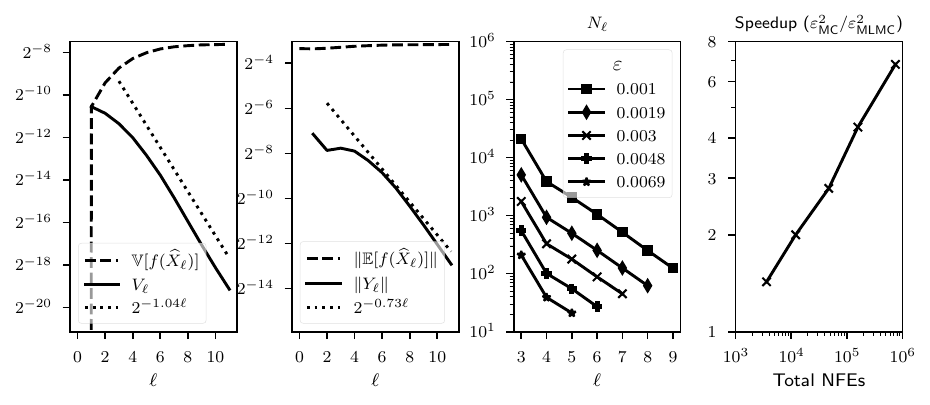}
    \caption{Results for the denoising (DN) experiment in \cref{sec:exp-denoising}}
    \label{fig:Denoising}
    \end{subfigure}
    \begin{subfigure}{0.8\textwidth}
    \centering
    \includegraphics[width=\textwidth]{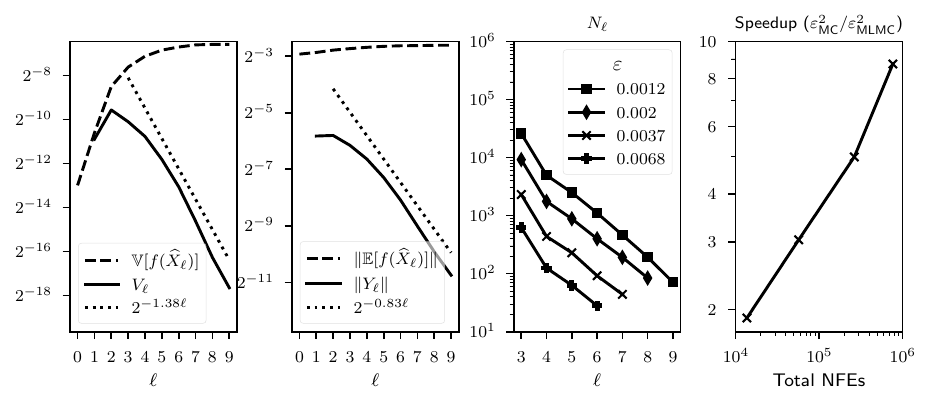}
    \caption{Results for the inpainting (IP) experiment in \cref{sec:exp-inpainting}}
    \label{fig:Inpainting}
\end{subfigure}
\caption{For each experiment we show, from left-to-right, the empirical
  variance estimate of $\mathbb{V}[f(\aX_{\ell})]$ and
  $V_{\ell}=\mathbb{V}[f(\aX_{\ell})-f(\aX_{\ell-1})]$, showing convergence of the
  latter as \(\ell\) increases with rate \(\beta\); the empirical mean estimates of
  $\lVert\mathbb{E}[f(\aX_{\ell})]\rVert$ and
  $\lVert\mathbb{E}[f(\aX_{\ell})-f(\aX_{\ell-1})]\rVert$, showing convergence of
  the latter as \(\ell\) increases with rate \(\alpha\); the number of required
  samples at each level to reach a particular error tolerance, showing that a
  smaller number of samples is required as \(\ell\) increases; and finally the
  MSE comparison between Monte Carlo and MLMC as a function of the total
  number of NFEs, showing that for same computational effort, MLMC achieves a
  smaller MSE. For inpainting and denoising, we have $T_0=M^{\ell_0}=2^3$ steps;
  For super-resolution, $T_0=M^{\ell_0}=2^5$. }
\end{figure}

We now estimate the marginal second moment of each image pixel to various precisions via MLMC and MC (see \cref{fig:Denoising}). MLMC substantially outperforms MC (e.g., to obtain a precision of 0.001, corresponding to $0.1\%$ of the pixel dynamic range, it is roughly $7\times$ more efficient to use MLMC rather than MC). The variance plot in the upper left shows that the variance increases as $\ell$ increases, in contrast to the super-resolution problem where we observed a large variance for small $\ell$. This stems from the fact that in this experiment the initial condition is fixed to the observation. In addition, the sampling scheme \cref{eq:DDPMUpdate} does not add noise in the final step, hence the variance is low when $\ell$ is small. 

\subsection{Experiment 3: Image inpainting}\label{sec:exp-inpainting}
We now consider a Bayesian image inpainting experiment, where one seeks to
perform inference on an unobserved region within the interior of an image. In
this experiment $y=(\mathbb{I}_n-\mathbb{M})x$, a form of
\cref{eq:InverseProblemNoisy} with $\mathcal{A}=(\mathbb{I}_n-\mathbb{M})$ and $\eta=0$.
$\mathbb{M}\in\mathbb{R}^{n\times n}$ is a binary diagonal matrix with diagonal entries set to
$1$ or $0$ to represent observed and unobserved image pixels respectively. In
our experiments, we consider that the unobserved pixels sit in a region in the
centre of the image, as illustrated in \cref{fig:ip}. We tackle this problem
by using the image-to-image Schr\"{o}dinger bridge ($\text{I}^2\text{SB}$)
model of \cite{Liu2023}, where $p_T(\cdot\vert
y)=\phi_n(\cdot;(\mathbb{I}_n-\mathbb{M})y,\mathbb{M})$ (see \cref{item:2,item:3} in
\cref{sec:DDPM}) and use the sampling scheme \cref{eq:DDPMUpdate} with a
mapping $s_{\theta}$ defined through a neural network. For the network, we use the
implementation\footnote{\url{https://github.com/NVlabs/I2SB}} from
\cite{Liu2023} which was trained for inpainting with a center mask on the
ImageNet 256×256 dataset via \cref{eq:LossFunctionDM}, with
$x_t^{(i)}=\mu_{t;0,T}(x_0^{(i)},y^{(i)})$ defined in \cref{eq:Sampling1}
deterministically (rather than drawn from \(\phi_{n}(\cdot;
\mu_{t;0,T},\sigma_{t;0,T}^2\mathbb{I}_{n})\)). It is worth noting that, unlike the
previous models which rely on stochastic mappings, $\text{I}^2\text{SB}$
adopts a deterministic push-forward approach that draws the unknown image
pixels from a standard normal distribution that is iteratively pushed forward
to the target posterior of interest by using \cref{eq:DDPMUpdate} with the
noise removed. This corresponds to the mean of the kernel $\rK_{t-1;0,t}(\cdot\vert
x_0,x_t)$ with $x_0$ estimated (implicitly) via the function $s_{\theta}$, so the
only source of stochasticity comes from the starting $x_T\sim\mathcal{N}((\mathbb{I}_{n}-
\mathbb{M}) y, \mathbb M)$. In the continuous limit, $\text{I}^2\text{SB}$
corresponds to the solution of an ``optimal transport ODE'' \cite{Liu2023},
and hence one still expects the discretisation error to decrease as more steps
are added, as is vital for the validity of the MLMC method. \cref{fig:ip}
shows an example of an image $x_0$ and the corresponding partial observation
$y$ that we seek to inpaint, together with a posterior samples as generated by
the considered $\text{I}^2\text{SB}$ model with \(10^{3}\) steps.

For this experiment, we again calculate the marginal second moment of each image pixel, but modify the norm used slightly, taking the Euclidean 2-norm of only the unobserved pixels in $x_0$ (since the other pixels will show no variation, being already determined in the observation $y$). Results are shown in \cref{fig:Inpainting}. Again, we observe that MLMC substantially outperforms MC (e.g., to obtain a precision of 0.0012, it is roughly $9\times$ more efficient to use MLMC rather than MC). Note that in this case $\beta>1$, so the cost dependence is slightly better than the $\beta=1$ case considered for the other two experiments, as to be expected from \cref{eq:MLMCAsymptoticCost}. The $\alpha,\beta$ for this experiment are also slightly larger than the other two experiments ($(0.8,1.4)$ versus $(0.7,1)$). This is due to the fact that the $\text{I}^2\text{SB}$ model relies on a deterministic mapping, unlike the models considered previously which used stochastic mappings. Experiments not shown here confirm that switching from stochastic to deterministic sampling leads to increased $\alpha,\beta$. Moreover, the lack of stochasticity, except in the initial condition, leads to small variance for small $\ell$.


\section{Conclusion}
This paper constitutes a proof of principle of the application of MLMC to
Bayesian computation via diffusion models. MLMC could also be applied, for
example: to the ensemble averaging performed in \cite{Price2023} for weather
prediction models; in combination with different discretisation levels of the
images ($16\times16, 32\times32$ etc.) \cite{Hagemann2023}; in combination with the
aforementioned strategies for reducing the cost of diffusion models (e.g.
pruning, distillation). Typically, the advantages of MLMC will be greater when
the underlying variance and uncertainty of the problem is greater - if more
NFEs are required, the advantage of MLMC over MC only grows. Therefore, MLMC
could be especially beneficial for scientific applications that rely on
large-scale foundational diffusion models, such as the new AURORA model of the
atmosphere \cite{bodnar2024aurora}. {For future work, having established
  that MLMC can deliver substantial reductions in computational cost for DMs,
  it would be natural to explore the combination of MLMC with DM distillation
  and quantization strategies. Moreover, MLMC could also be used to reduce
  variance in the training process for DMs, when fine-tuning a foundational DM
  or training it from scratch, as has been done for other models in, e.g.,
  \cite{zakariaei:mlmc-training}. 
}


\paragraph{Funding}
  A-L.H-A: UK Research and Innovation (UKRI) Engineering and Physical Sciences Research Council (EPSRC) through grant (EP/Y006143/1).
  M.P and K. Z: UKRI EPSRC through grants BLOOM (EP/V006134/1) and LEXCI (EP/W007681/1).
  L. S: Ministerio de Ciencia e Innovación (Spain) through project PID2022-136585NB-C21, MCIN/AEI/10.13039/501100011033/FEDER, UE.

\appendix
\section*{Supplement: Diffusion Models and SDEs}

\paragraph{DMs as SDE samplers}
One may also formulate DMs as numerical solvers for (reversed) SDEs \cite{Song2020}. We present the following in a more general unconditional context, in which the target distribution is not necessarily a posterior, and is simply denoted $p(\cdot)$.
The SDE in $\R^n$ (where $-A_t,B_t:[0,T]\mapsto\R_+$) where $W_t$ is a vector in $\R^n$ of $d$ independent Brownian motions $\left\{W_t^{(i)}\right\}_{i=1}^n$,
\begin{equation}\label{eq:dXt}
    dX_t=A_t X_t dt +B_tdW_t, \qquad X_0=x_0,
\end{equation}
has solutions conditioned on $x_0$ $X_t=\e^{\int^t_0 A_s ds}\left(x_0+\int_0^t\e^{-\int^s_0 A_u du}B_sdW_s\right)$, which are consequently distributed according to a Gaussian density $\K_{t;0}(\cdot\vert x_0)=\phi_n(\cdot;\mu_{t}(x_0),\Gamma_t\mathbb{I}_n)$. Indeed, in general one has
\begin{equation}\label{eq:SDEDDPM}
K_{t;s}(\cdot \vert x_s)=\phi_n\left(\cdot;\e^{\int_s^tA_udu}x_s,\left(\Gamma_t-\e^{2\int_s^tA_udu}\Gamma_s\right)\mathbb{I}_n\right).
\end{equation}
Denoting the distribution of solutions to the SDE at time $t$ as $p_t(\cdot)$, one may choose $A,B$ such that $\lim_{t\to\infty}\K_{t;0}(\cdot\vert x_0)=\phi_n(\cdot;0,\Gamma_t\mathbb{I}_n)$ \cite{Song2020}. Hence, with $x_0$ distributed according to the target $p(\cdot)$,
\begin{gather*}
    \lim_{t\to\infty}p_t(X)\propto\int_{x_0} \lim_{t\to\infty}\exp\left[-\Gamma_t^{-1}\left(X-\mu_t(x_0)\right)^2\right]p(x_0)dx_0=\exp\left[-\Gamma_t^{-1}X^2\right]\int_{x_0} p(x_0)dx_0=\phi_n(X;0,\Gamma_t\mathbb{I}_n).
\end{gather*}
And so by taking $T$ sufficiently large it is possible to reduce the error associated with the approximation $\phi_n(\cdot;0,\Gamma_t\mathbb{I}_n)\approx p_T(\cdot)$. 
The equation \cref{eq:dXt} has an associated reverse SDE (with $\tau=T-t$) \cite{Anderson1982}
\begin{equation}\label{eq:dYt}
    dX^r_{\tau}=\left[-A_{T-\tau} X^r_{\tau}+B_{T-\tau}^2\nabla\log(p_{T-\tau}(X^r_{\tau}))\right]d\tau+B_{T-\tau}dW_{\tau},
\end{equation}
whose solutions at $\tau=T$, if $X^r_0$ has probability density $p_T(\cdot)$, one can show, will give $X^r_T$ distributed according to $p(\cdot)$. Thus, in practice, by taking $A,B$ such that $X^r_0$ is drawn according to the density $\phi_n(\cdot)\approx p_T(\cdot)$ for fixed $T=1$ (finite time error), training a function $s_{\theta}(x,t)\approx \nabla\log p_{t}(x)$ (model error), discretising (discretisation error) and numerically solving the reverse SDE up to a time $\delta$ close to 0 (truncation error), one may generate approximate samples from the distribution $p(x_0)$ with the same four sources of error, and the same measure of cost (the number of $s_{\theta}$ evaluations) as for the discrete case. 

The so-called `score' $\nabla\log(p_{T-\tau}(\cdot))$, which encodes information about $p(\cdot)$, is not analytically available, and so must be approximated (parametrically) by a function $s_{\theta}(x,t)$. Such a function is easily trainable via a dataset of samples $\left\{x_0^{(i)}\right\}_{i=1}^N$ owing to the law of iterated expectation (with $\mathcal{T}=\{t^{(j)}\}_{k=1}^K$ a random index set of times $t^{(j)}\sim\mathcal{U}_{[0,T]}$)
\begin{align*}
\mathbb{E}_{x,t}[\lVert s_{\theta}(x,t)-\nabla\log p_t(x) \rVert^2]=\frac{1}{N\vert\mathcal{T}\vert}\approx \sum_{t\in \mathcal{T}}\sum_{i=1}^N \lVert s_{\theta}(x^{(i)}_t,t)+\Gamma_t^{-1}(x^{(i)}_t-\mu_t(x^{(i)}_0))\rVert^2
\end{align*}
since $\nabla\log K_{t;0}(x\vert x_0)=-\Gamma_t^{-1}(x-\mu_t(x_0))$ and $x^{(i)}_t$ has density $\phi_n(\cdot;\mu_t(x^{(i)}_0),\Gamma_t\mathbb{I}_n)$.


The resulting approximate reverse SDE must then be solved numerically, for example via the Euler-Maruyama scheme
\begin{equation}
    \label{eq:EulerMaruyama}X^r_{\tau_{n+1}}=X^r_{\tau_{n}}+h\left[-A_{T-\tau_{n}}X^r_{\tau_{n}}+B_{T-\tau_{n}}^2s_{\theta}(X^r_{\tau_n},T-\tau_n)\right]+\sqrt{h}B_{T-\tau_{n}}\xi,\qquad \xi\sim \mathcal{N}(0,1),
\end{equation}
introducing a third source of approximation error (numerical bias) proportional to the timestep discretisation $h=|\tau_{n+1}-\tau_n|$. Note however that one has a great deal of latitude in the choice of the numerical scheme used to solve \cref{eq:dYt}, as discussed below.
Typically $p(\cdot)$ will not be sufficiently well-behaved (e.g. it will be supported on a manifold of dimension $<n$) and so one solves up to $\tau=T-\delta$ (`truncation') to avoid numerical issues.

\begin{figure}
\centering
\begin{subfigure}{\textwidth}
\includegraphics[width=\textwidth]{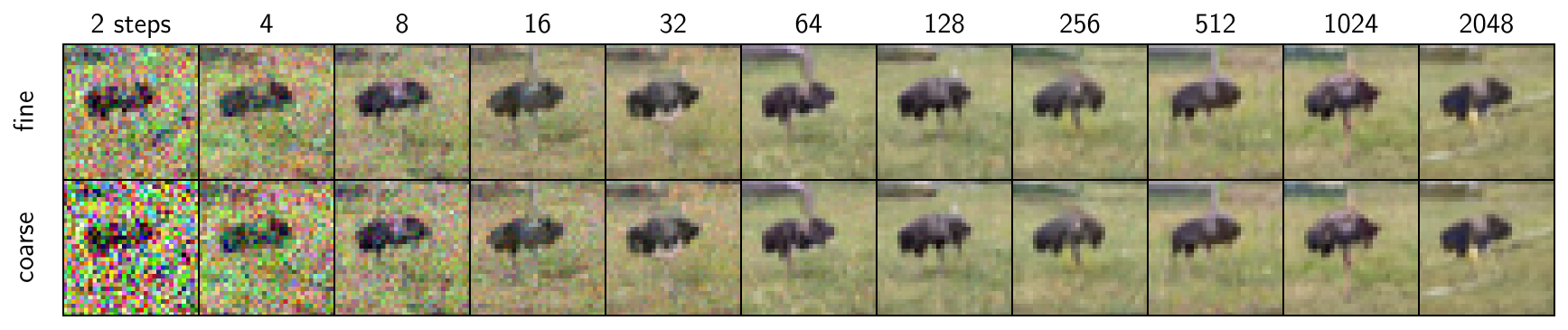}
\end{subfigure}
\\
\begin{subfigure}{\textwidth}
\includegraphics[width=\textwidth]{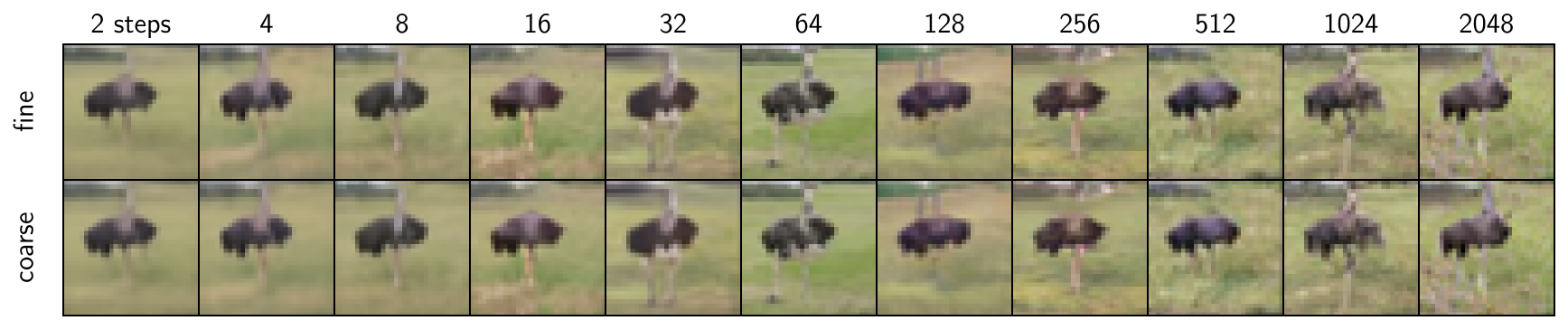}
\end{subfigure}
\caption{Above: Conditional denoised samples generated for Experiment 2 \cite{Song2020} using the EM scheme \cref{eq:EulerMaruyama} show instability for less than $\approx 2^5=32$ steps. Samples generated with fewer steps are unusable for the MLMC technique, since their variance does not fulfill the condition \cref{eq:Optimal_l0}. Below: Using the same network $s_{\theta}$ and SDE, but applying the DDIM1 numerical scheme \cref{eq:DDIM1} one is able to generate usable samples for MLMC with $2^2=4$ steps, which enables one to extract greater benefit from its usage.}
\label{fig:StabilityEM}
\end{figure}

\paragraph{Choice of Integrator}
In the case of DMs explicitly based on SDEs, one has greater latitude in the choice of the update step \cref{eq:DDPMUpdate}, which may be seen as a step of a numerical integrator \cite{Song2020}. One possibility is the Euler-Maruyama integrator \cref{eq:EulerMaruyama} - however this has poor stability properties, which manifest in an inability to generate realistic samples for large step sizes (small number of steps) (see \cref{fig:StabilityEM}). This is a problem for MLMC since one sees greater benefits from the multilevel nature of the technique, at a wider range of precisions, the more levels one can use.

However, discrete-time DMs of the type in \cref{eq:DDPMdecomp} are sometimes able to generate useful samples for $\sim20$ timesteps, and it has been shown that such models amount to exponential-type integrators combined with a reparametrised expression for the score function $s_{\theta}$ \cite{Zhang2022}. We summarise the derivation in the succeeding.

In \cite{Zhang2022}, one begins from the expression $\nabla\log \K_{s;0}(x_s\vert x_0)$ (since $\mu_s(x_0)=\e^{\int_0^s A_udu}x_0$ is linear)
\begin{equation*}
    \nabla\log \K_{s;0}(x_s\vert x_0)=-\Gamma_s^{-1}(x_s-\mu_sx_0),
\end{equation*}
which may be rearranged to give
$x_0=\mu_s^{-1}(x_s+\Gamma_s\nabla\log \K_{s;0}(x_s\vert x_0))$ and thus it holds for any two times $t,s$ that
\begin{equation*}
    \nabla\log \K_{t;0}(x_t\vert x_0)=-\Gamma_t^{-1}\left[x_t-\e^{\int_s^t A_udu}\left(x_s+\Gamma_s\nabla\log \K_{s;0}(x_s\vert x_0)\right)\right].
\end{equation*}
Then one may modify the RSDE \cref{eq:dYt} by replacing $\nabla\log p_t(x)\approx\nabla\log \K_{t;0}(x\vert x_0)$ and using the expression above. Denoting $A(T-\tau)=A^r_{\tau}$ etc. one has
\begin{align*}
        dX^r_{\tau}&=\left[-A^r_{\tau} X^r_{\tau}-\frac{(B^r_{\tau})^2}{\Gamma^r_{\tau}}\left[X^r_{\tau}-\e^{-\int_{\tau_s}^{\tau} A^r_udu}\left(X^r_{\tau_s}+\Gamma^r_{\tau_s}\nabla\log p^r_{\tau_s}(X^r_{\tau_s})\right)\right]\right]d\tau+B_{\tau}^rdW_{\tau}\\
        &=\left[-\left(A^r_{\tau}+\frac{(B^r_{\tau})^2}{\Gamma^r_{\tau}}\right)X^r_{\tau}+\frac{(B^r_{\tau})^2}{\Gamma^r_{\tau}}\e^{-\int_{\tau_s}^{\tau} A^r_udu}\left(X^r_{\tau_s}+\Gamma^r_{\tau_s}\nabla\log p^r_{\tau_s}(X^r_{\tau_s})\right)\right]d\tau+B_{\tau}^rdW_{\tau},
\end{align*}
where we must change the direction of integration\footnote{Since $\int_{T-\tau_s}^{T-\tau}A_udu=-\int_{\tau_s}^{\tau}A_{T-v}dv=-\int_{\tau_s}^{\tau}A^r_{v}dv$.}. This has exact solution between $[\tau_s=T-s,\tau_t=T-t]$ (replacing\footnote{And using that $d/dt\Gamma_t=B_t^2+2A_t\Gamma_t$ so that $d/d\tau\Gamma^r_{\tau}=-(B^r_{\tau})^2-2A^r_{\tau}\Gamma^r_{\tau}$.} $\nabla\log p^r_{\tau_s}(X^r_{\tau_s})\equiv f_{\tau_s}(X^r_{\tau_s})$)
\begin{gather*}
\begin{aligned}
        X^r_{\tau_t}&=\e^{-\int_{\tau_s}^{\tau_t}A^r_{\tau}+\frac{(B^r_{\tau})^2}{\Gamma^r_{\tau}}d\tau}\left[X^r_{\tau_s}+\int_{\tau_s}^{\tau_t}\frac{(B^r_{\tau})^2}{\Gamma^r_{\tau}}\e^{\int_{\tau_s}^{\tau} \frac{(B^r_{u})^2}{\Gamma^r_{u}}du}d\tau\left(X^r_{\tau_s}+\Gamma^r_{\tau_s}f_{\tau_s}(X^r_{\tau_s})\right)+\int_{\tau_s}^{\tau_t}B_{\tau}^r\e^{\int_{\tau_s}^{\tau}A^r_{u}+\frac{(B^r_{u})^2}{\Gamma^r_{u}}du}dW_{\tau}\right]\\
        &=\e^{-\int_{\tau_s}^{\tau_t}A^r_{\tau}d\tau}\left[X^r_{\tau_s}+(1-\e^{-\int_{\tau_s}^{\tau_t} \frac{(B^r_{\tau})^2}{\Gamma^r_{\tau}}d\tau})\Gamma^r_{\tau_s}f_{\tau_s}(X^r_{\tau_s})\right]+\int_{\tau_s}^{\tau_t}B_{\tau}^r\e^{-\int_{\tau}^{\tau_t}A^r_{u}+\frac{(B^r_{u})^2}{\Gamma^r_{u}}du}dW_{\tau}\\
        &=\e^{-\int_{\tau_s}^{\tau_t}A^r_{\tau}d\tau}X^r_{\tau_s}+\left(\e^{-\int_{\tau_s}^{\tau_t}A^r_{\tau}d\tau}\Gamma_{\tau_s}^r-\Gamma_{\tau_t}^r\e^{\int_{\tau_s}^{\tau_t} A^r_{\tau}d\tau}\right)f_{\tau_s}(X^r_{\tau_s})+\sqrt{\Gamma^r_{\tau_t}}\int_{\tau_s}^{\tau_t}\frac{B_{\tau}^r}{\sqrt{\Gamma^r_{\tau}}}\e^{-\int_{\tau}^{\tau_t}\frac{(B^r_{u})^2}{2\Gamma^r_{u}}du}dW_{\tau}.
\end{aligned}
\end{gather*}
This is then simulated numerically via the following scheme, called DDIM1 (with $s>t$)
\begin{equation}\label{eq:DDIM1}
\resizebox{1.1\hsize}{!}{$
    X^r_{\tau_t}=\e^{-\int_{t}^{s}A_{u}du}X^r_{\tau_s}+\left(\e^{-\int_{t}^{s}A_{u}du}\Gamma_{s}-\Gamma_{t}\e^{\int_{t}^{s} A_{u}du}\right)s_{\theta}(X^r_{\tau_s},s)+\sqrt{\Gamma_{t}\left(1-\Gamma_{t}\Gamma_{s}^{-1}\e^{2\int_{t}^{s}A_{u}du}\right)}\xi,\quad\xi\sim\mathcal{N}(0,1),
$}
\end{equation}
which is clearly equivalent to the discrete process \cref{eq:DDPMUpdate}.
To adapt the sampling to \cref{alg:ContinuousCoarseFine}, one needs to find the analogue of \cref{eq:variancecoarse}. One notes that
\begin{gather}
\begin{aligned}
\sqrt{\Gamma^r_{\tau_t}}\int_{\tau_s}^{\tau_t}&\frac{B_{\tau}^r}{\sqrt{\Gamma^r_{\tau}}}\e^{-\int_{\tau}^{\tau_t}\frac{(B^r_{u})^2}{2\Gamma^r_{u}}du}dW_{\tau}=\sqrt{\Gamma^r_{\tau_t}}\left[\int_{\tau_s}^{\tau_{t'}}\frac{B_{\tau}^r}{\sqrt{\Gamma^r_{\tau}}}\e^{-\int_{\tau}^{\tau_{t}}\frac{(B^r_{u})^2}{2\Gamma^r_{u}}du}dW_{\tau}+\int_{\tau_{t'}}^{\tau_t}\frac{B_{\tau}^r}{\sqrt{\Gamma^r_{\tau}}}\e^{-\int_{\tau}^{\tau_t}\frac{(B^r_{u})^2}{2\Gamma^r_{u}}du}dW_{\tau}\right]\\
&=\sqrt{\Gamma^r_{\tau_t}}\left[\e^{-\int_{\tau_{t'}}^{\tau_{t}}\frac{(B^r_{u})^2}{2\Gamma^r_{u}}du}\int_{\tau_s}^{\tau_{t'}}\frac{B_{\tau}^r}{\sqrt{\Gamma^r_{\tau}}}\e^{-\int_{\tau}^{\tau_{t'}}\frac{(B^r_{u})^2}{2\Gamma^r_{u}}du}dW_{\tau}+\int_{\tau_{t'}}^{\tau_t}\frac{B_{\tau}^r}{\sqrt{\Gamma^r_{\tau}}}\e^{-\int_{\tau}^{\tau_t}\frac{(B^r_{u})^2}{2\Gamma^r_{u}}du}dW_{\tau}\right]\\
&=\sqrt{\Gamma^r_{\tau_t}}\left[\e^{-\int_{\tau_{t'}}^{\tau_{t}}\frac{(B^r_{u})^2}{2\Gamma^r_{u}}du}\frac{\eta_{\tau_s,\tau_{t'}}}{\sqrt{\Gamma^r_{\tau_{t'}}}}+\frac{\eta_{\tau_{t'},\tau_{t}}}{\sqrt{\Gamma^r_{\tau_{t}}}}\right],
\end{aligned}
\end{gather}
where
$$
\eta_{\tau_{s},\tau_{t}}=\sqrt{\Gamma^r_{\tau_t}}\int_{\tau_s}^{\tau_t}\frac{B_{\tau}^r}{\sqrt{\Gamma^r_{\tau}}}\e^{-\int_{\tau}^{\tau_t}\frac{(B^r_{u})^2}{2\Gamma^r_{u}}du}dW_{\tau}
$$
is the noise added in a step from $\tau_{s}\to\tau_{t}$. Thus given two fine increments $\eta_{\tau_{s},\tau_{s-1}},\eta_{\tau_{s-1},\tau_{s-2}}$, one may generate the coarse increment $\eta_{\tau_{s},\tau_{s-2}}$ via
\begin{equation}
    \eta_{\tau_{s},\tau_{s-2}}=\sqrt{\Gamma^r_{\tau_{s-2}}}\left[\e^{-\int_{\tau_{s-1}}^{\tau_{s-2}}\frac{(B^r_{u})^2}{2\Gamma^r_{u}}du}\frac{\eta_{\tau_{s},\tau_{s-1}}}{\sqrt{\Gamma^r_{\tau_{s-1}}}}+\frac{\eta_{\tau_{s-1},\tau_{s-2}}}{\sqrt{\Gamma^r_{\tau_{s-2}}}}\right]=scaler(s)\eta_{\tau_{s},\tau_{s-1}}+\eta_{\tau_{s-1},\tau_{s-2}}.\label{eq:noisereweightingSDE}
\end{equation}
It can be seen that an iterative update of the form \cref{eq:variancecoarse} is also suitable to sum the fine increments for general $M$.

\end{document}